# The Ideation Bottleneck: Decomposing the Quality Gap Between AI-Generated and Human Economics Research

Li Ning

Tsinghua University

lining@sem.tsinghua.edu.cn

April 2026

**Abstract**

Autonomous AI systems can now generate complete economics research papers, but they substantially underperform human-authored publications in head-to-head comparisons. This paper decomposes the quality gap into two independent components: research idea quality and execution quality. Using a two-model ensemble of fine-tuned language models trained on publication decisions (Gong, Li, and Zhou, 2026) to evaluate idea quality and a comprehensive six-dimension rubric assessed by Gemini 3.1 Flash Lite — the same model family used as the APE tournament judge, ensuring methodological consistency — to evaluate execution quality, we analyze 953 economics papers — 912 AI-generated papers from the APE project and 41 human papers published in the American Economic Review and AEJ: Economic Policy. The idea quality gap is large (Cohen's d = 2.23, p < 0.001), with human papers achieving 47.1% mean ensemble exceptional probability versus 16.5% for AI. The execution quality gap is also significant but smaller (d = 0.90, p < 0.001), with human papers scoring 4.38/5.0 versus 3.84. Idea quality accounts for approximately 71% of the overall quality difference, with execution contributing 29%. The largest execution weakness is mechanism analysis depth (d = 1.43); no significant difference is found on robustness. We document that 74% of AI papers employ difference-in-differences, and only 7 AI papers (0.8%) surpass the median human paper on both idea and execution quality simultaneously. The primary bottleneck to competitive AI-generated economics research remains ideation.

# Introduction

The emergence of autonomous AI systems capable of generating complete research papers represents a fundamental shift in how scientific knowledge is produced (Lu et al., 2026; Korinek, 2023). Recent work has demonstrated that large language models can automate substantial portions of the research pipeline, from hypothesis generation to experimental execution (Si et al., 2024; Novy-Marx and Velikov, 2025). In economics specifically, the APE project (Autonomous Policy Evaluation), developed at the Social Catalyst Lab at the University of Zurich, has demonstrated that large language models can autonomously identify policy-relevant research questions, design identification strategies, collect and analyze data, and produce full research manuscripts without human intervention. Initial benchmarking of these AI-generated papers against human-authored publications from the American Economic Review and the American Economic Journal: Economic Policy reveals a substantial quality gap: in head-to-head comparisons evaluated by LLM judges, human papers win approximately 83 percent of matchups. Yet this aggregate comparison conceals a critical ambiguity. Research quality is not a single construct. It encompasses the quality of the underlying research idea — the originality and policy relevance of the question, the appropriateness of the chosen identification strategy, the strength of the proposed mechanism — and the quality of its execution — the rigor of the econometric implementation, the robustness of the empirical specification, the clarity of the exposition. Conflating these two dimensions prevents a precise diagnosis of where autonomous research systems succeed and where they fall short.

The distinction between ideation and implementation is not merely taxonomic. A brilliant idea poorly executed may lose to a mediocre idea brilliantly executed in a holistic judgment, yet the two cases call for entirely different remedies. For AI research systems, understanding which bottleneck binds has direct implications for system design: if AI struggles primarily with idea generation, then improvement efforts should focus on search processes, domain grounding, and novelty mechanisms; if AI struggles with execution, then attention should turn to econometric toolkits, code reliability, and output verification. The credibility revolution in empirical economics, as surveyed by Angrist and Pischke (2010), established precisely this kind of conceptual decomposition as productive: identification strategy design — a fundamentally idea-level choice about what variation to exploit — is analytically distinct from the econometric implementation that translates a chosen design into credible estimates. Our study adopts this same logic to decompose the AI-human research quality gap into its constituent parts.

To measure idea quality independently of execution, we draw on the fine-tuned language models developed by Gong, Li, and Zhou (2026) for evaluating economics research ideas. These supervised fine-tuned (SFT) models — language models further trained on domain-specific labeled data — were trained on thousands of editorial decisions and publication outcomes to predict the publication potential of standardized research idea descriptions. Each paper's research question and identification approach is compressed into a structured paragraph of 120–150 words following a fixed protocol that captures the phenomenon, gap, question, strategy, and contribution while deliberately excluding methods details, results, and writing quality. Applied to these compressed descriptions, the models output a probability distribution over four tiers of publication potential and a scalar exceptional probability score. Crucially, this measure is designed to assess the idea itself, independent of how well it was ultimately implemented in the paper. By applying the same evaluation instrument to both AI-generated and human-authored papers, we obtain a comparable idea-quality signal across the two populations.

To measure execution quality, we complement the SFT idea evaluation with a systematic rubric-based assessment using Gemini 3.1 Flash Lite, scoring each paper on six dimensions: causal identification strength, econometric sophistication, robustness and sensitivity analysis, data quality and appropriateness, mechanism analysis depth, and writing clarity and precision. Using the same model family as the APE tournament judge ensures that our execution assessments are methodologically consistent with the benchmark that established the original AI-human performance gap. The rubric draws on established standards from the recent methodological literature, including advances in difference-in-differences estimation with heterogeneous treatment effects (de Chaisemartin and D'Haultfoeuille, 2020; Goodman-Bacon, 2021; Callaway and Sant'Anna, 2021), coefficient stability bounds for assessing omitted variable bias (Oster, 2019), and cluster-robust and heteroskedasticity-robust inference standards (Abadie, Athey, Imbens, and Wooldridge, 2023). This rubric captures what a methodologically sophisticated referee would assess when evaluating not what was studied, but how rigorously it was studied.

Applying both evaluations to a corpus of 953 economics papers — 912 AI-generated papers from the APE project and 41 human-authored papers from the American Economic Review and AEJ: Economic Policy — we find that both dimensions show significant gaps, though of very different magnitudes. The idea quality gap, measured by a two-model SFT ensemble (GPT-4.1-nano-econ and Econ-30B), is large and highly significant: human papers achieve a mean ensemble exceptional probability of 47.1 percent against 16.5 percent for AI papers, with a standardized effect size of Cohen's $d = 2.23$ ($p < 0.001$). The execution quality gap is also statistically significant but smaller: human papers score 4.38 out of 5.0 against 3.84 for AI papers, with Cohen's $d = 0.90$ ($p < 0.001$). A variance decomposition attributes approximately 71 percent of the overall AI-human quality gap to differences in idea quality and 29 percent to differences in execution quality. Only 7 AI papers out of 912 (0.8 percent) surpass the median human paper on both dimensions simultaneously. Within execution, the mechanism analysis dimension shows the largest gap ($d = 1.43$), reflecting AI papers' tendency to rely on established causal architectures without probing boundary conditions or constructing rich theoretical accounts of why effects arise. No significant difference is found on robustness and sensitivity analysis ($d = 0.08$, $p = 0.827$).

Beyond the main decomposition, we document an additional pattern with implications for understanding the idea quality gap: 74 percent of AI-generated papers employ difference-in-differences as their primary identification strategy, compared with a substantially more diverse methodological toolkit among human authors. This methodological concentration is likely not incidental. The APE pipeline generates research by searching for natural experiments and policy discontinuities, a process that structurally favors DiD-amenable quasi-experiments. This narrowness in identification approach may constrain the originality of the resulting research questions, since the most novel economics questions often require either novel data sources or novel methodological frameworks rather than the application of a standard estimator to a newly identified policy variation. The remainder of this paper proceeds as follows. Section 2 describes the data and the APE corpus. Section 3 details the idea quality evaluation methodology. Section 4 describes the execution quality rubric and scoring procedure. Section 5 presents the main results and decomposition. Section 6 discusses implications for the design of autonomous research systems and the future of AI in economics. Section 7 concludes.

## 2. Data and Methods

### 2.1 Data

The empirical analysis draws on two primary sources. The first is the Autonomous Policy Evaluation (APE) project (ape.socialcatalystlab.org), an initiative housed at the Social Catalyst Lab at the University of Zurich that has produced 912 economics research papers generated entirely by large language models. These papers span the complete research pipeline – each system autonomously identifies a policy-relevant question, designs an identification strategy, collects data, executes the empirical analysis, and writes up the results. The APE project evaluates these papers via a tournament system using TrueSkill Bayesian skill ratings (Herbrich, Minka, and Graepel, 2007) — a ranking algorithm that models each paper's quality as a Gaussian distribution updated after each pairwise comparison — with Google's Gemini 3.1 Flash Lite as the pairwise judge comparing papers head-to-head on five dimensions: identification strategy, novelty, policy relevance, execution quality, and scope. In this tournament, human papers win 82.9% of head-to-head matchups.

The second source is a set of recently published human benchmark papers. The APE project includes 43 human papers drawn from two leading economics journals: 9 articles from the American Economic Review and 34 from the American Economic Journal: Economic Policy, both from the 2025–2026 publication period. Two of the 34 AEJ: Economic Policy papers were inaccessible due to paywall restrictions, yielding a final human sample of 41 papers. Together, the AI and human samples comprise 953 papers for analysis. The AI papers were obtained from the APE GitHub repository in LaTeX source format; human papers were downloaded as PDFs from links provided on the APE project website and converted to plain text for processing.

An important scope distinction underlies the entire analysis. The APE tournament evaluates the complete paper – including data collection, empirical execution, robustness checks, and writing quality. The present study decomposes this overall quality judgment into two independent components: the quality of the underlying research idea, and the quality of its execution. These measure fundamentally different constructs, and their joint analysis allows us to attribute the AI-human performance gap to its proximate sources.

### 2.2 Idea Quality Evaluation

To isolate research idea quality from execution quality, we adopt the standardized idea description framework developed in Gong, Li, and Zhou (2026, arXiv:2603.16659). Each paper's research idea is compressed into a structured paragraph of 120–150 words — termed an "idea summary" in the original protocol — that encapsulates five elements in a fixed order: the empirical phenomenon under study, the gap in existing knowledge, the central research question, the proposed identification strategy, and the claimed theoretical or policy contribution. This compressed representation deliberately excludes data availability, empirical results, robustness evidence, and writing quality, forcing evaluators to assess only the intellectual substance of the idea itself.

Extraction proceeds from each paper's front matter — the abstract and introduction only — to further limit the influence of execution-specific content. Each paper's front matter was processed through a large language model (Qwen 3.6 Plus, accessed via the OpenRouter API[1]) using the extraction prompt specified in Gong, Li, and Zhou (2026; see Appendix B for the full prompt text). The extraction protocol instructs the model to represent the paper's framing faithfully and without embellishment: it must not infer methodological sophistication beyond what the authors state, must not interpolate results or findings not present in the front matter, and must preserve the authors' own language for the research question and identification strategy where possible. This objectivity constraint is critical for ensuring that extraction quality does not itself introduce systematic differences between AI and human papers.

Extracted descriptions are evaluated by a two-model ensemble of fine-tuned classifiers: GPT-4.1-nano-econ (a fine-tuned version of OpenAI's GPT-4.1-nano) and Econ-30B (a locally hosted 30-billion-parameter model fine-tuned on the same training protocol). Both models were trained on thousands of historical publication decisions in the social sciences, with training labels derived from the institutional prestige tier of the publishing venue (Gong, Li, and Zhou, 2026). Each model assesses ideas on two dimensions – originality and usefulness – and classifies them into four tiers: Exceptional (field-defining originality combined with broad policy or welfare relevance, corresponding to AER, Econometrica, QJE, JPE, and Review of Economic Studies standards), Strong (clear contribution in either identification strategy or policy relevance, corresponding to the AEJ series, RAND Journal of Economics, or Journal of Development Economics), Fair (incremental extensions of known findings in predictable directions), and Limited (weak on both originality and policy relevance). Each model is constrained to output a single token (its tier classification) with full log-probability information over the vocabulary.[2] The log-probability scores associated with the four tier tokens are converted to a probability distribution via softmax normalization (exponentiating each score and dividing by the sum of exponentiated scores). The ensemble exceptional probability is the average of the two models' exceptional probabilities, providing a more robust idea quality measure than either model alone.

The central methodological advantage of this approach is isolation. By feeding each evaluator only a standardized idea description rather than a full paper, the SFT models' judgments cannot be contaminated by data quality, empirical execution, or writing clarity. A mediocre idea executed brilliantly may rank highly in the APE tournament; it will not receive a high exceptional probability from our ensemble unless the idea itself is genuinely original and useful.

---

[1] OpenRouter is an API routing service that provides unified access to language models from multiple providers (openrouter.ai).

[2] Specifically: max_tokens=1, temperature=0, top_logprobs=20. This forces the model to commit to a single-token classification while revealing its confidence across all four tiers through the log-probability distribution. The two models exhibit different calibration profiles — GPT-4.1-nano produces higher absolute exceptional probabilities than Econ-30B — but agree strongly on the relative ordering of papers. Averaging stabilizes the estimates and reduces model-specific biases.

## 2.3 Execution Quality Evaluation

To evaluate the execution dimension, we developed a comprehensive scoring rubric through systematic AI-assisted review of methodological guidance from across the empirical economics literature. This review drew on referee guidelines published by leading journals including the American Economic Review, the Quarterly Journal of Economics, and Econometrica, as well as foundational methodological texts (Angrist and Pischke, 2009, 2010) and recent advances in causal inference methodology. The resulting rubric incorporates current best practices on heterogeneous treatment effects in two-way fixed effects designs (de Chaisemartin and D'Haultfoeuille, 2020), decomposition of staggered difference-in-differences estimators (Goodman-Bacon, 2021), group-time average treatment effects (Callaway and Sant'Anna, 2021), the power of pre-testing procedures (Roth, 2022), coefficient stability under omitted variable bias (Oster, 2019), and the clustering of standard errors in applied work (Abadie, Athey, Imbens, and Wooldridge, 2023).

The rubric evaluates each paper along six dimensions, each weighted to reflect its centrality to causal credibility in contemporary applied economics: Identification Strategy (25%), Econometric Methodology (20%), Robustness and Sensitivity (20%), Data Quality (15%), Mechanism and External Validity (10%), and Writing and Presentation (10%). Each dimension is scored on a 1–5 scale using explicit, criterion-referenced rubric language. For the Identification Strategy dimension, scoring criteria are tailored to the specific method employed by the paper – difference-in-differences, instrumental variables, regression discontinuity, randomized control trial, or other – with method-specific benchmarks reflecting what constitutes adequate, good, or exceptional practice under each design. This method-conditional scoring avoids penalizing papers for the inherent limitations of their identification approach and instead evaluates how well the paper executes within the design it has chosen.

Evaluations were conducted by Google's Gemini 3.1 Flash Lite, a lightweight large language model optimized for high-throughput analytical tasks, following the LLM-as-judge paradigm established by Zheng et al. (2023). This choice is methodologically consistent with the APE tournament itself, which also uses Gemini 3.1 Flash Lite as its pairwise judge, ensuring that our execution assessments draw on the same model family used to establish the original AI-human performance gap. While LLM evaluators are subject to known biases including position sensitivity and verbosity preference (Zheng et al., 2023), these biases are mitigated in our design by evaluating each paper independently (no pairwise comparison) and constraining output to structured JSON scores rather than free-form judgments. Each paper was evaluated in an independent session using the full paper text – methods section, data description, results, and robustness checks – rather than the abstract alone. Fresh context was initialized for each paper to eliminate any cross-paper contamination in the model's assessments.

The composite execution score is the weighted average of the six dimension scores, yielding a continuous quality measure on a 1–5 scale.

## 2.4 Methodological Diversity

As a complement to the quantitative quality measures, we characterize the methodological composition of each corpus by classifying papers according to their primary identification strategy. These classifications were produced as a byproduct of the execution evaluation: the rubric requires the evaluator to identify the empirical design before applying method-specific scoring criteria, yielding a categorization of each paper's approach as difference-in-differences, instrumental variables, regression discontinuity, randomized control trial, structural estimation, or a mixed or unclassified strategy.

The AI corpus exhibits striking methodological concentration. Difference-in-differences accounts for 74% of AI-generated papers, compared to 24% of human papers in the benchmark sample. Human authors employ a considerably more diverse toolkit: instrumental variables (17%), randomized control trials (10%), mixed or multi-design approaches (20%), and structural estimation (5%). This asymmetry in methodological diversity likely reflects both the composition of the training data on which the AI systems were optimized and the relative ease of generating textually coherent papers that follow the standard difference-in-differences template. As we discuss in Section 4, this concentration has implications for both the idea-quality and execution-quality scores: the identification strategy dimension of the execution rubric rewards papers that select the most credible design for their question, and a corpus dominated by difference-in-differences will tend to receive similar identification scores regardless of whether a different design might have been preferable.

# 3. Results

## 3.1 Idea Quality

Research idea quality is evaluated by a two-model SFT ensemble comprising GPT-4.1-nano-econ and Econ-30B, both fine-tuned on publication decisions. The individual models exhibit different calibration profiles: GPT-4.1-nano assigns AI papers a mean exceptional probability of 0.309 and human papers 0.605, while Econ-30B is substantially more conservative, assigning 0.021 and 0.338 respectively. This calibration difference is itself informative — the two models agree on the direction and rough magnitude of the gap but disagree on the absolute level, reflecting known variation in fine-tuned model confidence. The ensemble average stabilizes these estimates, yielding mean exceptional probabilities of 0.165 for AI papers and 0.471 for human papers.

Using the tier predictions from GPT-4.1-nano (the more calibrated of the two models), AI papers distribute as 22.9% Exceptional, 65.1% Strong, 10.4% Fair, and 1.5% Limited; human papers as 73.2% Exceptional, 26.8% Strong, and none rated Fair or

Limited. A chi-square test confirms the distributional difference ($\chi^2$ = 53.48, df = 3, p < 0.001). Figure 1a displays the tier distributions side by side; Figure 1b shows the full exceptional probability distributions.

The ensemble exceptional probability gap of 0.306 is highly significant by a Mann-Whitney U test (z = -7.61, p < 0.001), with a large standardized effect size (Cohen's d = 2.23). The effect size is amplified relative to either individual model because the ensemble compresses the absolute scale while preserving the relative gap. Despite this aggregate difference, the AI distribution is not uniformly weak. Several AI papers achieve ensemble exceptional probabilities approaching or exceeding the human median (0.472), though this remains rare. Only 7 AI papers (0.8 percent) simultaneously exceed the median human paper on both idea quality and execution quality — a finding we examine in detail below.

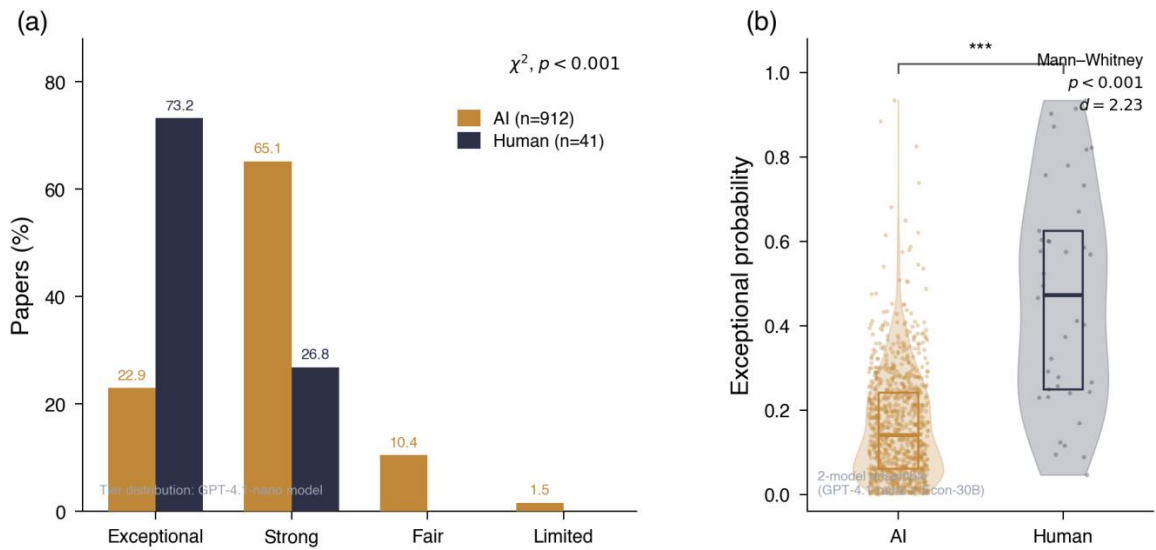

*Figure 1. Idea quality comparison.* (a) Tier distribution of AI-generated (n = 912) and human-authored (n = 41) papers based on GPT-4.1-nano tier predictions. (b) Distribution of two-model ensemble exceptional probability scores (GPT-4.1-nano + Econ-30B averaged). Cohen's d = 2.23.

## 3.2 Execution Quality

Gemini 3.1 Flash Lite's holistic ratings across six execution dimensions reveal a statistically significant execution gap alongside the idea quality gap. Composite scores are 3.84 out of 5 for AI papers and 4.38 for human papers. A Mann-Whitney test confirms this difference is statistically significant (p < 0.001, Cohen's d = 0.90), though the effect size is meaningfully smaller than the idea quality gap (Figure 2a).

The dimension-level results, summarized in Table 1 and visualized in the radar chart (Figure 2b), reveal important within-composite heterogeneity. The Mechanism and External Validity dimension shows the largest gap (AI: 3.32, Human: 4.20; d = 1.43, p < 0.001), reflecting AI papers' tendency to rely on established causal architectures without probing boundary conditions or constructing rich theoretical accounts of why effects arise. Data Quality shows the next largest gap (AI: 4.03, Human: 4.85; d = 1.05, p < 0.001), followed by Identification Strategy (AI: 3.69, Human: 4.41; d = 0.81, p < 0.001) and Econometric Methodology (AI: 3.71, Human: 4.24; d = 0.74, p < 0.001).

**Table 1: Execution Quality by Dimension**

| Dimension | AI Mean | Human Mean | Cohen's d | p-value |
| --- | --- | --- | --- | --- |
| Identification Strategy | 3.69 | 4.41 | 0.81 | < 0.001 *** |
| Econometric Methodology | 3.71 | 4.24 | 0.74 | < 0.001 *** |
| Robustness & Sensitivity | 3.91 | 3.98 | 0.08 | 0.827 ns |
| Data Quality | 4.03 | 4.85 | 1.05 | < 0.001 *** |
| Mechanism & External Validity | 3.32 | 4.20 | 1.43 | < 0.001 *** |
| Writing & Presentation | 4.58 | 4.88 | 0.55 | < 0.001 *** |

*Note:  **p < 0.01,**  p < 0.001, ns = not significant.*

One dimension shows no statistically significant difference: Robustness and Sensitivity analysis (AI: 3.91, Human: 3.98; d = 0.08, p = 0.827). AI and human papers perform equivalently on this procedural dimension, suggesting that AI systems do apply standard robustness protocols with similar thoroughness. Writing and Presentation shows a significant but relatively modest gap (AI: 4.58, Human: 4.88; d = 0.55, p < 0.001). The mechanism analysis dimension, where AI lags by the widest margin, likely reflects AI papers' tendency to apply causal architectures mechanically without probing boundary conditions, constructing theoretical accounts of why effects arise, or systematically addressing external validity.

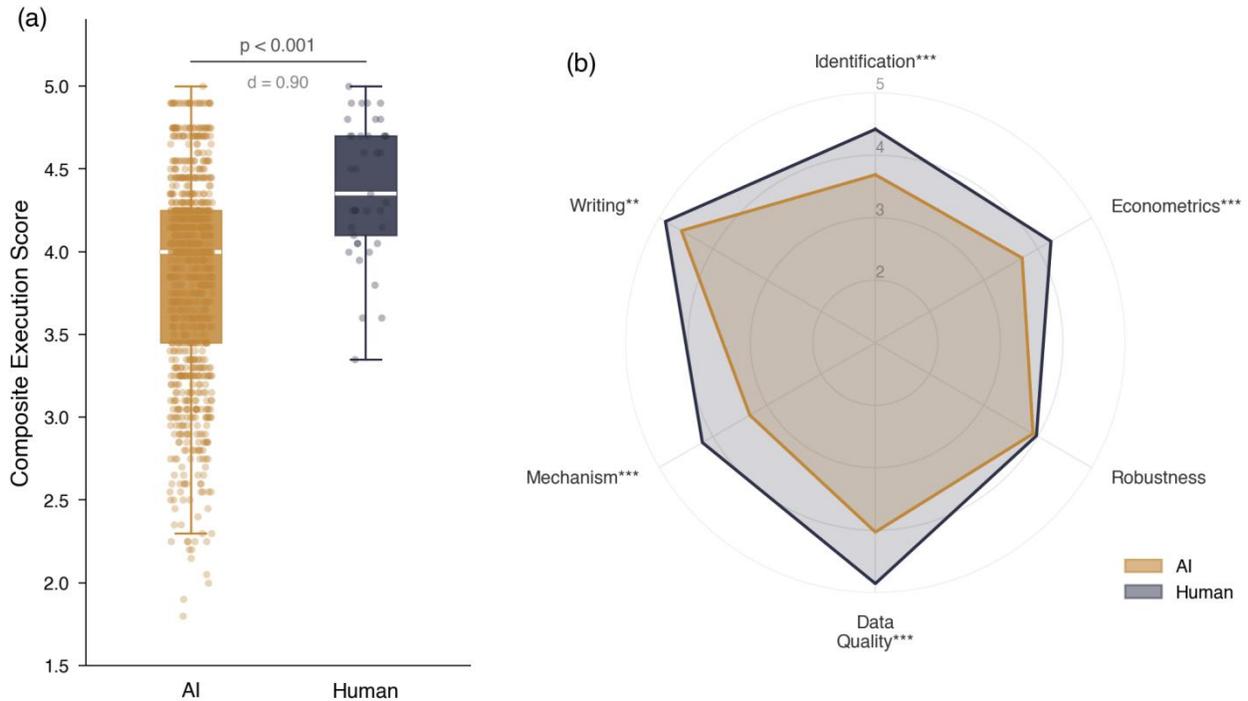

*Figure 2. Execution quality comparison (Gemini 3.1 Flash Lite). (a) Composite execution scores (p < 0.001, d = 0.90). (b) Radar chart of six execution dimensions. No significant difference on robustness; humans lead on mechanism analysis (d = 1.43) and data quality (d = 1.05).*

## 3.3 Decomposing the Quality Gap

The simultaneous availability of idea-quality and execution-quality measures for the same papers permits a variance decomposition of the total human-AI performance differential. With the two-model SFT ensemble, the idea-quality gap corresponds to Cohen's d = 2.23, while the execution-quality gap corresponds to d = 0.90. The idea-quality effect is approximately 2.5 times larger than the execution-quality effect (Figure 3a). A proportional attribution assigns approximately 71% of the total quality gap to idea quality and 29% to execution quality. Both gaps are statistically significant, but idea quality is clearly the dominant component.

This decomposition is reinforced by the low correlation between the two measures. Across all 953 papers, the Pearson correlation between ensemble exceptional probability and execution composite score is r = 0.098, indicating that the two assessments capture largely independent attributes of research quality (Figure 3b). A paper can pose a weak research question while executing it meticulously, or vice versa, and the measures reflect this independence. This discriminant validity supports the interpretability of the decomposition.

To assess how many AI papers are genuinely competitive with human research across both dimensions, we compare each AI paper against the median human paper on both idea quality and execution quality simultaneously. Only 7 of 912 AI papers (0.8 percent) exceed the human median on both the ensemble exceptional probability and the Gemini execution composite (Figure 4). A more lenient threshold — exceeding the 25th percentile of human papers on both dimensions — yields 76 AI papers (8.3 percent). At the floor — exceeding the minimum human score on both dimensions — 585 AI papers (64.1 percent) qualify. This funnel analysis reveals that while the majority of AI papers exceed the weakest human papers, truly competitive AI research that matches or surpasses typical human quality remains rare.

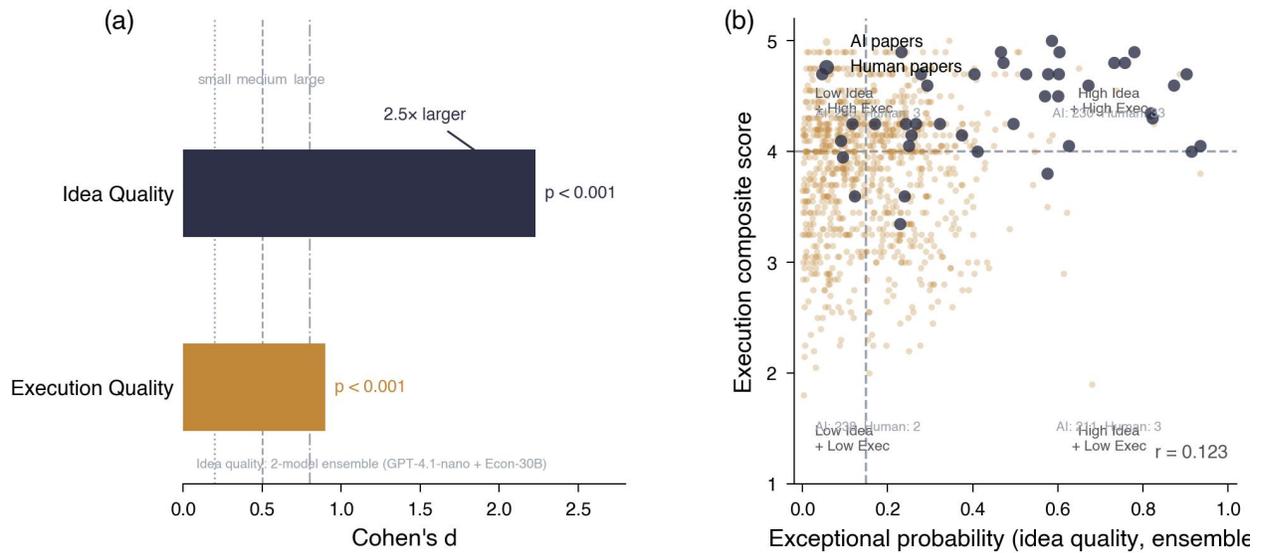

*Figure 3. Decomposing the quality gap.* (a) Standardized effect sizes: idea quality (d = 2.23) is 2.5 times larger than execution quality (d = 0.90). (b) Scatter plot of ensemble idea quality versus execution quality (r = 0.12), confirming independent constructs.

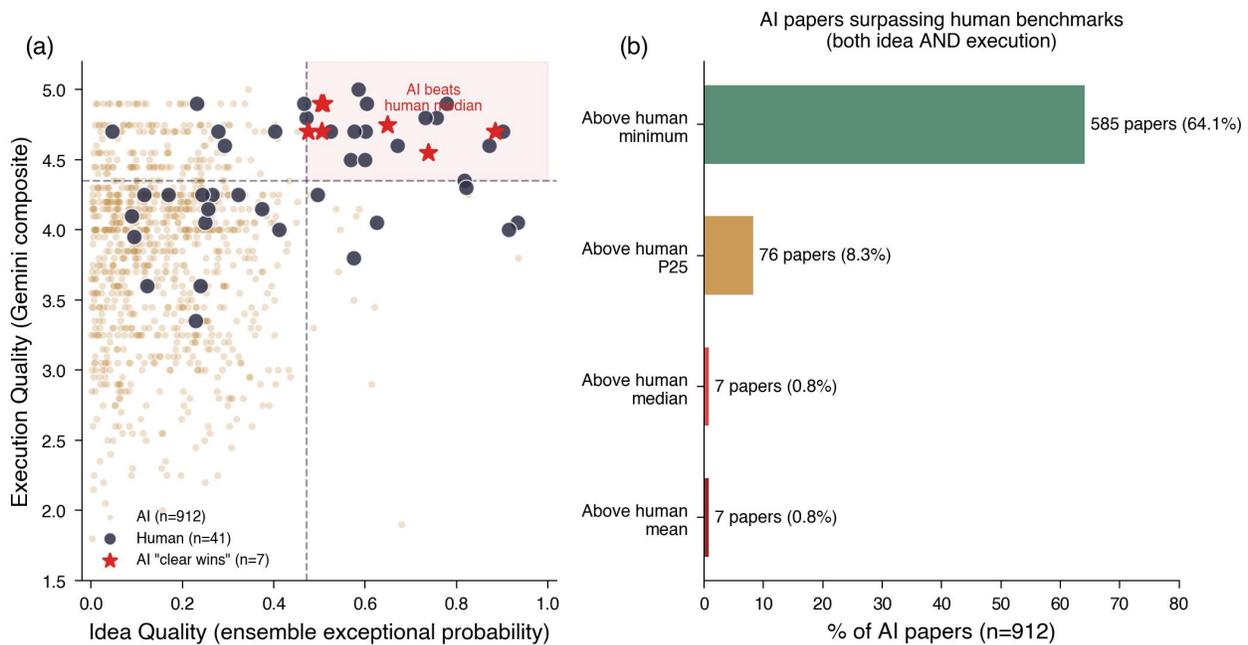

*Figure 4. AI papers that surpass human benchmarks.* *(a) Scatter plot with human median thresholds (dashed lines). Red stars mark the 7 AI papers (0.8%) that exceed the median human paper on both idea and execution quality. (b) Funnel analysis showing progressively stricter thresholds.*

## 3.4 Methodological Diversity

AI-generated papers exhibit a striking concentration in their choice of empirical strategy. Difference-in-differences designs account for 74% of AI papers, followed distantly by regression discontinuity (11%), mixed methods (6%), instrumental variables (5%), descriptive analysis (2%), and randomized controlled trials (0.2%). Human papers in the sample draw on a considerably broader toolkit: difference-in-differences (24%), instrumental variables (17%), mixed methods (20%), regression discontinuity (15%), randomized controlled trials (10%), descriptive analysis (10%), and structural estimation (5%). Figure 5a visualizes this contrast. The AI corpus thus exhibits a mono-methodological character, with nearly three-quarters of papers clustering around a single identification strategy. This homogeneity suggests that AI ideation systems may have internalized a stylized template of modern applied microeconometrics—DiD as the default—rather than developing sensitivity to the contextual features of a research question that determine which identification strategy is appropriate.

## 3.5 Cross-Ranking Analysis

A natural validation exercise is to examine how our quality measures correlate with APE tournament rankings, which are determined by LLM judges comparing full papers on dimensions including data, execution, and argumentation. Spearman rank correlations between APE tournament standing and each quality measure are shown in Figure 5b. The execution composite ($\rho = 0.398$) predicts APE rank more strongly than does the idea exceptional probability ($\rho = 0.216$), and a combined measure yields $\rho = 0.422$, a modest improvement over execution alone. The stronger predictive power of execution quality is consistent with the nature of the tournament: judges assessing complete papers are exposed to the full research product, including data construction, robustness tables, and exposition, all of which align more directly with the execution dimensions than with the standalone quality of the research question. Among the individual execution sub-dimensions, data quality and identification strategy show the strongest correlations with tournament rank, reinforcing their importance in holistic paper evaluation.

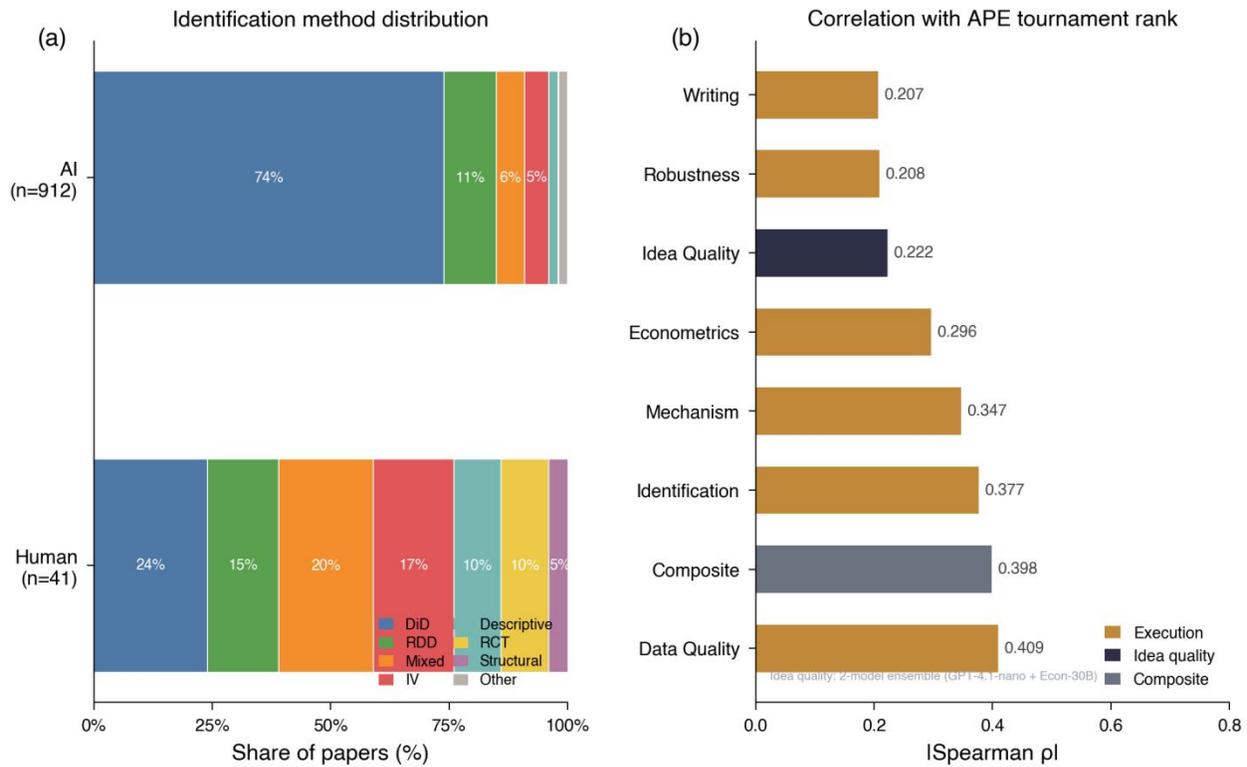

*Figure 5. Methodological diversity and tournament predictors.* (a) Distribution of identification strategies — AI papers concentrated in difference-in-differences (74%). (b) Spearman rank correlations between quality sub-dimensions and APE tournament ranking.

# 4. Discussion and Conclusion

## 4.1 What the Decomposition Reveals

The central finding of this paper is that both idea quality and execution quality contribute meaningfully to the AI-human research gap, but idea quality remains the dominant source of underperformance. The execution gap is statistically significant ($p < 0.001$, $d = 0.90$), but the idea quality gap is substantially larger ($d = 2.23$, $p < 0.001$). A variance decomposition attributes approximately 71% of the total performance difference to idea quality and 29% to execution quality. The binding constraint on automated economics research is primarily creative ideation, but execution also falls meaningfully short of human standards. This finding echoes Si et al. (2024), who found in a large-scale human study that LLM-generated research ideas in NLP were judged more novel but weaker on feasibility than expert ideas — a pattern consistent with the idea that AI systems can generate plausible-sounding ideas but struggle with the deeper originality that domain experts produce.

The implication requires some reframing of the standard narrative. When economists discuss the prospects of AI-generated research, debate often centers on whether AI can handle the technical apparatus of modern empirical economics — the instrumentation strategies, the robustness checks, the data construction. Our results suggest that AI is not yet fully competitive on execution either, though it comes closer to human standards there than on ideation. The specific weaknesses are informative: AI papers lag most severely on mechanism analysis ($d = 1.43$) and data quality ($d = 1.05$), suggesting that the limitations are not uniformly distributed across execution dimensions. AI systems can correctly implement difference-in-differences designs and achieve robustness testing that is statistically indistinguishable from human work ($d = 0.08$, $p = 0.827$), but they struggle to probe why effects arise, construct rich theoretical accounts of mechanisms, and leverage high-quality administrative data in the same way human economists can.

The robustness and sensitivity finding is worth underscoring. AI and human papers perform equivalently on this procedural dimension, suggesting that AI systems do apply standard robustness protocols with similar thoroughness. This is a genuine competency of LLM-generated research: systematic enumeration of alternative specifications is a task that AI systems appear to execute as rigorously as human authors working under time and cognitive constraints. The larger challenge lies in the qualitative dimensions of research quality — mechanism construction and data sourcing — where pattern matching on existing papers cannot substitute for the substantive domain judgment that shapes how human economists design and ground their empirical work.

Finally, the near-zero correlation between idea quality and execution quality ($r = 0.098$) provides methodological validation for the decomposition itself. These are not two faces of a single latent quality dimension but genuinely independent constructs. A paper can have a strong research question and mediocre implementation, or a weak question executed with technical rigor. This independence means that closing the gap requires targeted improvements on both fronts: ideation mechanisms capable of generating genuinely novel questions, and execution capabilities that can construct richer mechanisms and access higher-quality data.

## 4.2 The Methodological Narrowness Problem

A striking feature of the AI paper corpus is its concentration on a single identification strategy: 74% of AI-generated papers rely on difference-in-differences designs. By comparison, the human benchmark draws on a substantially broader methodological repertoire spanning instrumental variables, regression discontinuity, structural estimation, and randomized controlled trials. This concentration is not accidental — it likely reflects both the prevalence of DiD in LLM training corpora and the relative tractability of DiD framing for generative models.

What makes this methodological narrowness consequential is its recursive effect on ideation. Research questions are not generated in a vacuum; they are shaped by the methods available to answer them. If an AI system's effective methodological repertoire is largely limited to DiD designs, it can only generate questions amenable to DiD identification — questions about the effects of discrete policy changes on panel data. The vast space of research questions best addressed through IV, structural modeling, or experimental variation becomes systematically underrepresented. In this sense, the methodological narrowness may partially cause the idea quality gap rather than merely reflecting it. Expanding the breadth of identification strategies that LLMs can competently deploy would not only improve execution diversity but would also expand the space of original research questions that AI systems can credibly propose.

This observation points toward a specific direction for improving AI research generation: training and prompting strategies that explicitly reward methodological diversity, and fine-tuning on corpora that represent the full spectrum of identification approaches used in top journals. Progress on methodological breadth may yield disproportionate returns to idea quality by unlocking question-types that are currently inaccessible to systems constrained to a narrow set of empirical templates.

## 4.3 Limitations

Several limitations bear on the interpretation of these findings. While the idea quality evaluation uses a two-model SFT ensemble (GPT-4.1-nano-econ and Econ-30B), both models were fine-tuned primarily on management and organizational behavior publication decisions with prompts adapted for economics. The two models exhibit substantially different calibration — Econ-30B assigns much lower absolute exceptional probabilities than GPT-4.1-nano — and the ensemble average, while more robust than either model alone, inherits the biases of its constituents. The execution evaluation relies on a single LLM evaluator — Gemini 3.1 Flash Lite — rather than human expert referees; while the rubric is grounded in published assessment standards and the choice of Gemini ensures consistency with the APE tournament methodology, systematic differences between LLM and human judgment cannot be ruled out. The human benchmark is also small, comprising 41 papers from two journals (AER and AEJ:Policy); a broader sample spanning QJE, JPE, ReStud, and Econometrica would strengthen the comparison and reduce the influence of journal-specific norms on the results. There is additionally a model mismatch in the idea extraction pipeline: rq_with_context strings were extracted using Qwen 3.6 Plus while the SFT evaluator was trained on extractions produced by Qwen3-235B, a discrepancy that could introduce noise into the idea quality scores. Finally, because AI-generated papers are produced as integrated wholes, the execution quality of a paper may leave traces in how its research question is framed, partially confounding the decomposition; the near-zero correlation between dimensions suggests this contamination is modest, but it cannot be entirely excluded.

## 4.4 Conclusion

The primary frontier of automated economics research is ideation, though execution also falls meaningfully short of human standards. Idea quality accounts for approximately 71% of the AI-human performance gap and execution for 29%; the bottleneck is not purely in either dimension but is substantially more concentrated in the creative act of identifying research questions than in the technical act of executing them. Human economists retain their largest comparative advantage in generating questions that are simultaneously original and consequential, and in constructing mechanism analyses and leveraging high-quality data in ways that require genuine domain judgment. Only 7 of 912 AI papers (0.8 percent) surpass the median human paper on both idea and execution quality simultaneously, illustrating how rare it remains for autonomous systems to produce research that is genuinely competitive across the board. The path toward competitive AI-generated economics research therefore requires progress on two fronts: expanding the methodological repertoire and novelty mechanisms that govern ideation, and improving the depth of mechanism analysis and the quality of data sourcing in execution. As AI-generated research becomes more prevalent across the social sciences, understanding this decomposition helps clarify where human expertise remains most valuable and where collaboration between human and machine researchers is likely to be most productive.

# Appendix

## Appendix A: Idea Quality Evaluation Prompt

The following system prompt was used by the fine-tuned GPT-4.1-nano-econ model (Gong, Li, and Zhou, 2026) to evaluate research idea quality. The model receives this prompt as a system message, followed by the standardized idea description as the user message. It outputs a single token with log-probability information enabled, and the probability distribution over the four tier tokens is extracted via softmax normalization (see Section 2.2 for details).

> **ROLE**: You are an expert evaluator of economics research ideas. Your task is to evaluate from a senior economist's perspective: be direct and critical, give clear judgments based on originality and usefulness to classify research ideas into appropriate publication potential tiers.

**TASK**: Read a paragraph describing an economics research idea and classify it into one of four publication potential tiers. Your classification should be based on two key dimensions: originality and usefulness. Output ONLY the tier notation with NO explanation or reasoning.

**EVALUATION CRITERIA**

*Originality* reflects whether the research idea addresses a genuinely open question, resolves an empirical puzzle, or challenges prior findings in a substantive way. Original research provides a new identification strategy, a novel causal mechanism, a previously unavailable empirical fact, or a theoretical framework that yields non-obvious predictions. The key question is whether the idea provides substantive new insight that changes how economists understand a phenomenon – not merely whether it applies known methods to a slightly different context.

*Usefulness* reflects whether the research idea offers policy-relevant evidence, welfare implications, or insights into important market, institutional, or behavioral mechanisms. Useful research tackles pressing economic or social challenges with broad implications – informing policy design, advancing economic theory in ways that reshape subsequent work, or establishing new empirical facts of wide interest.

**CLASSIFICATION TIERS**

*Exceptional*: Research that demonstrates both strong originality and strong usefulness. Field-defining contributions – a credible new identification strategy or theoretical innovation combined with broad policy or welfare relevance. Suitable for AER, Econometrica, QJE, JPE, ReStud.

*Strong*: Clear contribution in either identification/originality or policy relevance, with the other dimension adequately developed. Strong potential for leading field journals (AEJ series, RAND, JDE, JOLE, JIE, JPubE, JET).

*Fair*: Incremental contributions, extending known findings in predictable directions or applying established methods to modestly new settings. Mid-level field journals (BPEA, JEH, JPop).

*Limited*: Weak on both originality and policy relevance. Replicate well-known results, address narrow scope, or lack credible identification. Lower-tier or regional journals.

**OUTPUT FORMAT**: Respond with EXACTLY ONE of: Exceptional, Strong, Fair, or Limited.

# Appendix B: Research Idea Extraction Prompt (SM4)

Each paper's research idea was extracted into the standardized idea description format using the extraction prompt from Gong, Li, and Zhou (2026). The extraction model

(Qwen 3.6 Plus) receives the paper's abstract and introduction as input and produces a standardized 120–150 word paragraph. The prompt enforces objectivity: the model must represent the paper's framing faithfully without embellishment, added sophistication, or inferred contributions.

> **ROLE**: You are an objective research paper analyzer. Your task is to extract and present research questions and core elements from academic papers WITHOUT interpretation, embellishment, or improvement.
>
> **CRITICAL PRINCIPLE – OBJECTIVITY OVER PERSUASIVENESS**: Present the paper EXACTLY as written by the authors. Do NOT add theoretical sophistication if it is not there. Do NOT create compelling hooks if the original lacks them. Do NOT infer contributions beyond what authors explicitly state. Do NOT improve weak framing – describe it as presented. If the idea seems underdeveloped in the original, your summary should reflect that.
>
> **OUTPUT**: Generate ONLY the RQ_WITH_CONTEXT version (120–150 words, 1 paragraph): the phenomenon or problem (1–2 sentences), what is missing or unclear in existing research (2–3 sentences), the research question (1–2 sentences), the approach or identification strategy (1 sentence emphasizing the empirical strategy such as RCT, diff-in-diff, IV, RDD, or structural model), and the key claimed contribution (1 sentence).
>
> **EXTRACTION RULES**: Focus on the abstract, introduction, and theoretical development sections only. Use the authors' exact terminology for key constructs. Preserve the certainty level. Do NOT add theoretical connections, persuasive hooks, or inferred contributions not explicitly stated by the authors.

## Appendix C: Execution Quality Evaluation Rubric

The execution evaluation rubric was developed through systematic AI-assisted literature review, drawing on 30+ academic sources including referee guidelines from AER, QJE, and Econometrica; methodological textbooks (Angrist and Pischke, 2009, 2010); and recent advances in causal inference. Each paper was evaluated by Gemini 3.1 Flash Lite via the OpenRouter API, receiving the full paper text and the following system prompt. This choice is methodologically consistent with the APE tournament, which also uses Gemini 3.1 Flash Lite as its pairwise judge. Each evaluation was conducted in an independent session with fresh context to prevent cross-paper contamination.

The rubric scores six dimensions on a 1–5 scale:

**Dimension 1: Identification Strategy (25%).** Evaluates how convincingly the paper establishes causal claims, with method-specific criteria. For difference-in-differences: parallel trends evidence, staggered treatment handling per de Chaisemartin and D'Haultfoeuille (2020) and Callaway and Sant'Anna (2021), pre-trend testing with Roth (2023) caveats. For instrumental variables: exclusion restriction plausibility, first-stage F-statistic per Staiger and Stock (1997), reduced-form estimates, weak instrument diagnostics. For regression discontinuity: McCrary (2008) density test, bandwidth sensitivity per Calonico, Cattaneo, and Titiunik, covariate smoothness. Score 5 requires state-of-the-art identification with all assumptions explicit and defended. Score 1 indicates no clear identification strategy with assumptions unstated.

**Dimension 2: Econometric Methodology (20%).** Evaluates whether the estimation approach is appropriate and correctly implemented: standard errors appropriate with clustering justified by design per Abadie, Athey, Imbens, and Wooldridge (2023); multiple hypothesis testing corrections applied (Bonferroni, Holm, Benjamini-Hochberg, Romano-Wolf); coefficient stability bounds per Oster (2019); flexible functional form tested.

**Dimension 3: Robustness and Sensitivity (20%).** Evaluates how thoroughly results are stress-tested: placebo and falsification tests, alternative specifications, bandwidth and window sensitivity, subsample analysis and leave-one-out, specification curve or multiverse analysis, theory-driven heterogeneity analysis with pre-specified versus exploratory clearly labeled.

**Dimension 4: Data Quality (15%).** Evaluates the empirical foundation: data source quality (administrative data ranked highest, followed by experimental, survey, and convenience samples), sample size and statistical power, representativeness and generalizability, variable measurement (direct measures versus proxies), missing data and attrition handling, descriptive statistics completeness.

**Dimension 5: Mechanism and External Validity (10%).** Evaluates whether the paper goes beyond reduced-form effects: mediation analysis, heterogeneity patterns consistent with theoretical mechanism, ruling out alternative channels, LATE versus ATE distinction for instrumental variables, explicit external validity discussion with context dependence acknowledged.

**Dimension 6: Writing, Presentation, and Transparency (10%).** Evaluates whether the paper is well-crafted and transparent: clear structure with contribution upfront and framework before results, tables with standard errors in parentheses and dependent variable mean reported, economic magnitude discussed rather than only statistical significance, literature positioning with clear marginal contribution, limitations discussed with potential bias direction acknowledged, code and data availability.

The composite score is the weighted average: 0.25 x D1 + 0.20 x D2 + 0.20 x D3 + 0.15 x D4 + 0.10 x D5 + 0.10 x D6.

## Appendix D: APE Tournament Ranking Methodology

The APE tournament ranking is determined through pairwise comparison. Each pair of papers is evaluated twice by Gemini 3.1 Flash Lite (a non-Anthropic model selected to avoid self-preference bias), with the paper presentation order swapped between the two evaluations to control for position bias. The judge reads the complete paper – text, figures, tables, and formatting – and selects a winner with a confidence level (high, medium, or low), accompanied by structured reasoning organized around identification strategy, robustness, and policy relevance.

The judge rewards papers demonstrating novel questions that challenge conventional wisdom, rigorous identification including honest null results, and transparent engagement with limitations. It penalizes weak identification, failed validation tests or violated assumptions, and shallow analysis lacking robustness checks. The final match result aggregates the two position-swapped judgments: if both agree, the agreed-upon paper wins; if they disagree, the match is recorded as a tie.

Paper rankings accumulate through TrueSkill Bayesian ratings (Herbrich, Minka, and Graepel, 2007), a system originally developed for multiplayer game ranking that models each player's skill as a Gaussian distribution with mean (mu) and uncertainty (sigma). After each match, both the winning and losing paper's skill distributions are updated. The conservative rating (mu minus a multiple of sigma) determines the leaderboard position, penalizing papers with fewer matches and therefore greater rating uncertainty.

A representative judge reasoning excerpt from a human-versus-AI matchup illustrates the evaluation style:

> "Paper B employs a sophisticated spatial regression discontinuity design on a massive, high-quality administrative dataset (the universe of French property transactions), providing a highly credible causal estimate of the 'stigma' effect of place-based policy. While Paper A is a well-executed field study […]"

The judge's reasoning typically addresses three dimensions: (1) identification strategy and data quality, (2) robustness and empirical rigor, and (3) policy relevance and contribution. These unstructured text-based judgments do not produce separate numerical sub-scores; the winner is determined holistically.